\documentclass[apj]{emulateapj}

\shorttitle{instability due to gas-dust friction}
\shortauthors{M. Shadmehri}

\begin{document}

\title{Analysis of the instability due to gas-dust friction in protoplanetary discs}

\author{Mohsen Shadmehri}
\affil{Department of Physics, Faculty of Science, Golestan University, Gorgan 49138-15739, Iran \\Research Institute for Astronomy and Astrophysics of Maragha (RIAAM), Maragha, Iran, P. O. Box: 55134-441\\ m.shadmehri@gu.ac.ir}

\begin{abstract}
We study stability of a dust layer in a gaseous disc subject to the linear axisymmetric perturbations. Instead of considering single-size particles, however, the population of dust particles is  assumed to consist of two grain species. Dust grains exchange momentum with the gas via the drag force and their self-gravity is also considered.  We show that the presence of two grain sizes can increase the efficiency of the linear growth of drag-driven instability in the protoplanetary discs. A second dust phase with a small mass, comparing to the first dust phase, would reduce the growth timescale even by a factor of two or  more especially when its coupling to the gas is weak.  It means that once a certain amount of large dust particles form, even though it is much smaller than that of small dust particles, the dust layer becomes more unstable and dust clumping are accelerated. Thus,  presence of dust particles with various sizes must be considered in studies of dust clumping in protoplanetary discs where both large and small dust grains are present. 
\end{abstract}


\section{Introduction}
Planets are thought to born in accretion discs consisting of gas and dust around young stars, though formation mechanisms are still under intense  debate \citep[e.g.,][]{rafikov2005,matzner2005,chiang2010,armit}. These protoplanetary discs (PPDs) are complex systems, mostly  because of the diversity of physical processes that may affect their structure. A PPD is modeled as a multi-region system depending on its physical properties  like temperature, density, level of ionization, strength of the magnetic field and chemical composition. For example,  dominant sources of ionization such as  central radiation and cosmic rays can ionize the inner and outer  parts of a PPD, though between these two regions, at midplane of the disc, a non-ionized region is formed that none of the mentioned sources of ionization is able to ionize the gas \citep{deadzone}. This magnetically inactive or quiescent region is called dead zone which is sandwiched by active layers near the top and  bottom surfaces of  the disc.

Presence of dust grains not only affect thermal and radiation processes and even ionization level in a PPD but also understanding their dynamics is a vital step to gain more physical insight into formation of rocky planets  and  core of giant gaseous planets.  Various mechanisms have been proposed to operate at each phase of the planet  formation processes depending on the size of  dust particles. As for the dynamics of the dust grains, however, their momentum exchange with the gas component via the drag force plays an important role. This interaction largely depends on the level of dust-gas coupling which can be measured by the stopping time normalized by the orbital frequency, i.e. $\tau = t_{\rm stop} \Omega $ where $\Omega $ is the Keplerian rotational velocity. This dimensionless stopping time $\tau$ increases with the particle size.  When  coupling is strong (i.e., $\tau \ll 1$), as it is for small grains, motion of the particles is significantly modified because of their interactions with the gas component. 

While the gas component of a PPD  is subject to the gradient of the pressure and the central gravitational force, a dust grain only experiences gravity of the star and so, dust particles are rotating at full Keplerian velocity and the gas rotational profile is sub-Keplerian. In other words, orbital motions of the gas and dust grains are not the same. Actually this movement of dust grains through the gas component is the main driving mechanism of the so-called {\it streaming}  instability which leads to clumping of dust particles \citep[e.g.,][]{youdin2005,youdin2007,Jac,Laibe}. Although this instability is a promising route to planetesimal formation, its efficiency significantly reduces if the stopping time becomes less than unity for well-coupled particles.

Thus,  for clumping of particles with very small stopping times, however, it seems another mechanism is needed.  \cite{gold} presented a pioneering work of classical gravitational instability in dusty discs to explain how planetesimals are formed. \cite{cora} also studied formation of planetesimals in the sedimentating dust component of a protoplanetary disc by considering the interaction between dust particles  and gas component. \cite{sekiya83} performed the linear analyses of the gravitational instability of the incompressible gas. A mechanism which is actually driven by the drag force has  been proposed by many authors  and is known as {\it secular gravitational instability} \citep[e.g.,][]{goodman2001,Cuzzi,Mich,youdin2011}.  In the absence of dust particles, stability of a gaseous disc is determined via the well-known Toomre parameter, i.e. $Q\equiv c_{\rm s}\Omega / \pi G\Sigma$ where $c_{\rm s}$ is the sound speed and $\Sigma$ is the surface density. According to this criterion, as long as Toomre parameter stays at values larger than one, a disc is gravitational stable and once this condition is violated, small amplitude perturbations lead to fragmentation of the disc, though survival of the fragments strongly  depends on the cooling rate \citep[e.g.,][]{gammie2001}. Because of neglecting dissipative process, the standard Toomre condition is a dissipationless formation mechanism. In this regard, secular gravitational instability (SGI) is the dissipative version of Toomre analysis for two-component discs, in which the drag force is explicitly considered. An important feature of SGI is its ability to operate even in PPDs with a Toomre parameter larger than one.  Clumping of particles due to SGI is unconditional, though its growth rate highly depends on the properties of the disc like its metallicity which is defined as the ratio of dust and gas densities.   

Since the total mass of dust particles constitutes no more than one percent of the disc mass, drag force affects dynamics of dust grains much stronger than the gas component. For this reason, most of the previous linear studies of SGI assume that motion of the particles occurs through a fixed gaseous background. Considering dust-gas feedback leads to new results especially at long wavelengths as has been shown just recently by \cite{Taka}. They studied  drag-driven instability by modeling a mixture of dust and gas components in a shearing sheet system of coordinates \citep{gold}. \cite{Taka} found that not only the instability is no longer unconditional, but also in contrast to the prediction of the standard SGI studies, the system is stable at long wavelengths.

But population of dust particles in PPDs exhibits a wide range of sizes. This important aspect of the system has been neglected in all previous analytical studies of SGI in PPDs to our knowledge. Considering the vital role of the drag force in triggering the instability and the strong dependence of this force on the dust size, we think, it would be important to analyze the drag-driven instability in a dust layer with multiple sizes. This constitutes the main subject of the present study. In the next section, our main assumptions and equations are presented. We then obtain linearized equations and analyze resulting dispersion relation in sections 3 and 4.

\section{General Formulation}
We consider a PPD around a central star with mass $M$ as a system consisting of gas and dust components with  momentum exchanges via the drag force. It is assumed that the disc is thin so that the motion of both gas and dust fluids are in the plane of the disc. Our analysis is done in a local shearing box system \citep{gold}. We construct a rotating system on the neighborhood of a point $(r,\theta)=(r_0 , \Omega t)$, where the Keplerian angular velocity at this point is denoted by $\Omega$, i.e. $\Omega=\sqrt{GM/r_{0}^3}$. In this rotating system the local radial and azimuthal coordinated are $(x,y)=(r-r_0 , r_0 (\theta - \Omega t))$. 

Thus, basic equations for the gas component in this local system are
\begin{equation}
\frac{\partial \Sigma}{\partial t}+{\nabla}.(\Sigma\mathbf{u})=0 ,
\end{equation}
\begin{displaymath}
\Sigma(\frac{\partial\mathbf{u}}{\partial t}+\mathbf{u}.\nabla\mathbf{u}+2\mathbf{\Omega}\times\mathbf{u}-\Omega^2r)=
\end{displaymath}
\begin{equation}
-\Sigma \nabla(\Phi-\frac{GM}{r})-c_s^2\nabla\Sigma+\frac{\Sigma_{d1}(\mathbf{v_{\rm 1}-u)}}{t_{\rm stop,1}}+\frac{\Sigma_{d2}(\mathbf{v_{\rm 2}-u)}}{t_{\rm stop,2}},
\end{equation}
where $\Sigma$, $\bf{u}$ and $c_s$ are surface density, velocity and the sound speed of gas, respectively.  Moreover, $\Sigma_{d1}$ and $\Sigma_{d2}$ are surface density of the first and the second dust phases, respectively. We also assumed that the gas component is isothermal.  Moreover, ${\bf v}_1$ and ${\bf v}_2$ are velocity of dust particles with stopping times  $t_{\rm stop, 1}$ and $t_{\rm stop, 2}$, respectively. 

Also, the basic equations for the   dust phase one are written as
\begin{equation}
\frac{\partial\Sigma_{d1}}{\partial t}+\nabla.(\Sigma_{d1}\mathbf{v}_{1})=D_{1}\nabla^2\Sigma_{d1} , 
\end{equation}
\begin{displaymath}
\Sigma_{d1}(\frac{\partial\mathbf{v}_{1}}{\partial t}+\mathbf{v}_{1}.\nabla\mathbf{v}_{1}+2\mathbf{\Omega}\times\mathbf{v}_{1}-\Omega^2r)=
\end{displaymath}
\begin{equation}
-\Sigma_{d1} \nabla(\Phi-\frac{GM}{r})+\frac{\Sigma_{d1}(\mathbf{u-v_{\rm 1})}}{t_{\rm stop,1}},
\end{equation}
and for the  dust phase two, the basic equations are
\begin{equation}
\frac{\partial\Sigma_{d2}}{\partial t}+\nabla.(\Sigma_{d2}\mathbf{v}_{2})=D_{2}\nabla^2\Sigma_{d2} , 
\end{equation}
\begin{displaymath}
\Sigma_{d2}(\frac{\partial\mathbf{v}_{2}}{\partial t}+\mathbf{v}_{2}.\nabla\mathbf{v}_{2}+2\mathbf{\Omega}\times\mathbf{v}_{2}-\Omega^2r)=
\end{displaymath}
\begin{equation}
-\Sigma_{d2} \nabla(\Phi-\frac{GM}{r})+\frac{\Sigma_{d2}(\mathbf{u-v_{\rm 2})}}{t_{\rm stop,2}},
\end{equation}
where $D_1$ and $D_2$ are the diffusivities of the first and second phases of dust grains  because of the gas turbulence. Note that $\Phi$ is the gravitational potential due to both gas and dust fluids. Thus, the Poisson equation becomes
\begin{equation}
\nabla^2\Phi=4 \pi G(\Sigma+\Sigma_{d1}+\Sigma_{d2})\delta(z).
\end{equation}

 Although our formulation is based on previous studies \citep[e.g.,][]{Taka}, the equation of continuity with the diffusion term is not used commonly. In fact, the above hydrodynamical equations are  derived from the Boltzmann equation (or Fokker-Planck equation). This approach leads to the viscosity term in equation of motion but not lead to the diffusion term in equation of continuity. On the other hand, the turbulent stirring causes the velocity dispersion of dust particles \citep[e.g.,][]{youdin2007}. Therefore, it seems that the viscosity and pressure terms should be included in equation of motion instead of the diffusion term in equation of continuity if we consider equation of motion as we do here. This problematic formulation have been adopted in some of the previous works without validation or the explanation. We do not address these aspects of the problem here, but it deserves further work. 

 The diffusion of dust particles due to stochastic forcing by gas turbulence has been studied by many authors \citep[e.g.,][]{cuzzi93,youdin2007}. According to equation (36) of \cite{youdin2007}, the radial particle diffusion coefficients $D_1$ and $D_2$ are written as
\begin{equation}
D_1 = \frac{1+\tau_1 + 4 \tau_1^2}{(1+\tau_1^2)^2} D_{g},
\end{equation}
\begin{equation}
D_2 = \frac{1+\tau_2 + 4 \tau_2^2}{(1+\tau_2^2)^2} D_{g},
\end{equation}
where $\tau_1$ and $\tau_2$ are the nondimensional stopping times for the first and the second dust grain species, i.e. $t_{\rm stop,1} \Omega = \tau_1$ and $t_{\rm stop,2} \Omega = \tau_2 $. Moreover,  $D_g$ is the strength of turbulent diffusion in the gas. This quantity is defined via an $\alpha$ parameter:
\begin{equation}
D_{g}=\alpha c_{s}^2 \Omega^{-1},
\end{equation}
 where $\alpha$ is a dimensionless measure of turbulent intensity. \cite{youdin2011} extensively discussed that the diffusive parameter $\alpha$ with values from $10^{-10}$ to $10^{-6}$ is much smaller than a similar, but not equivalent, parameter in accretion disc studies.
We note that  although the first and the second dust  particle species do not interact directly with each other, they are still indirectly coupled because  of their interactions with the gas.
\section{Linear Perturbations}
 In order to present the linearized equations, we have to specify the initial configuration of the system. The initial surface density of the gas $\Sigma_0$ is assumed to be uniform and time-independent. Ratios of the dust densities  to the gas density are introduced as $\epsilon_{1}=\Sigma_{d1,0}/\Sigma_{0}$ and   $\epsilon_{2}=\Sigma_{d2,0}/\Sigma_{0}$ which are also constant. In the rotating system, components of the gas and dust initial velocities are ${\bf u}_0 =(-3/2) \Omega x {\bf j}$, ${\bf v}_{1,0}=(-3/2) \Omega x {\bf j}$ and ${\bf v}_{2,0}=(-3/2) \Omega x {\bf j}$, where ${\bf j}$ is unit vector in the y direction. We can now apply linear perturbations for all physical quantities as $\chi =\chi_0 +\delta \chi$ where $|\delta \chi | \ll |\chi_0 |$ and $\delta \chi$ is proportional to $\exp ({\rm i}kx-{\rm i}\omega t)$. Here, $k$ is the wavenumber of the perturbations and $\omega$ is  the frequency. Then, linearized equations become
\begin{equation}
-i\omega \delta\Sigma + ik\Sigma_{0} \delta u_{x}=0,
\end{equation}
\begin{displaymath}
-i\omega \delta u_{x} - 2\Omega \delta u_{y} = -c_{\rm s}^2 \frac{ik \delta\Sigma}{\Sigma_0}-ik\delta\Phi + \frac{\epsilon_{1} (\delta v_{1,x} - \delta u_{x})}{t_{\rm stop, 1}} 
\end{displaymath}
\begin{equation}
+ \frac{\epsilon_{2} (\delta v_{2,x} - \delta u_{x})}{t_{\rm stop, 2}},
\end{equation}
\begin{equation}
-i\omega \delta u_{y} + \frac{\Omega}{2} \delta u_{x} =\frac{\epsilon_{1}(\delta v_{1,y}-\delta u_{y})}{t_{\rm stop, 1}} + \frac{\epsilon_{2}(\delta v_{2,y}-\delta u_{y})}{t_{\rm stop, 2}},
\end{equation}
\begin{equation}
-i\omega \delta\Sigma_{d1}+ ik\epsilon_{1} \Sigma_{0} \delta v_{1,x}=-D_{1} k^2 \delta\Sigma_{d1},
\end{equation}
\begin{equation}
-i\omega \delta v_{1,x} - 2\Omega \delta v_{1,y} = -ik \delta\Phi + \frac{\delta u_{x} - \delta v_{1,x}}{t_{\rm stop, 1}},
\end{equation}
\begin{equation}
-i\omega \delta v_{1,y} + \frac{\Omega}{2} \delta v_{1,x} = \frac{\delta u_{y} - \delta v_{1,y}}{t_{\rm stop, 1}},
\end{equation}
\begin{equation}
-i\omega \delta\Sigma_{d2}+ ik\epsilon_{2} \Sigma_{0} \delta v_{2,x}=-D_{2} k^2 \delta\Sigma_{d2},
\end{equation}
\begin{equation}
-i\omega \delta v_{2,x} - 2\Omega \delta v_{2,y} = -ik \delta\Phi + \frac{\delta u_{x} - \delta v_{2,x}}{t_{\rm stop, 2}},
\end{equation}
\begin{equation}
-i\omega \delta v_{2,y} + \frac{\Omega}{2} \delta v_{2,x} = \frac{\delta u_{y} - \delta v_{2,y}}{t_{\rm stop, 2}},
\end{equation}
\begin{equation}\label{eq:Perturb-P}
\delta \Phi = -\frac{2\pi G}{|k|} (\frac{\delta\Sigma}{1+kH} + \frac{\delta\Sigma_{d1}}{1+kH_{d1}} + \frac{\delta\Sigma_{d2}}{1+kH_{d2}}).
\end{equation}

 Since our analysis is based on thin disc approximation, the factor $(1+kH)^{-1}$ appears in the perturbed Poisson equation (\ref{eq:Perturb-P}) due to the finite thickness of the disc \citep{van,shu84,romeo}. Here, $H$ is the thickness of the gaseous disc. In the absence of this correction factor the thin approximation breaks down and the effect of self-gravity can be overestimated if the wavelength of perturbations is not long. Note that the thickness correction factors for dust components, i.e. $(1+kH_{d1})^{-1}$ and $(1+kH_{d2})^{-1}$ are applied independently to each of them. The particle scale height for the first and the second dust phases are determined as $H_{d1} = \sqrt{\alpha/\tau_1} H$ and $H_{d2}=\sqrt{\alpha/\tau_2} H$ \citep{youdin2007}.

Based on the above equations, dispersion relation for the instability is obtained. The resulting dispersion relation obtained is tediously long and of limited interest, and so, we will not be reproduced it here for clarity. But analysis of its roots,  especially unstable modes, is performed in the next section. Obviously, the instability occurs if ${\rm Re}[-{\rm i}\omega]$ becomes positive.

\section{Analysis}
Our input parameters are the dimensionless stopping times $\tau_1$ and $\tau_2$, Toomre parameter $Q$ and the disc metallicities $\epsilon_{1}$ and $\epsilon_{2}$. We also define dimensionless diffusion coefficients $\xi_1 = D_1 /(c_{\rm s}^2 \Omega^{-1})$ and $\xi_2 = D_2 /(c_{\rm s}^2 \Omega^{-1})$. Thus, 
\begin{equation}
\xi_1 = \alpha \frac{1+\tau_1 + 4 \tau_1^2}{(1+\tau_1^2)^2},
\end{equation}
\begin{equation}
\xi_2 = \alpha \frac{1+\tau_2 + 4 \tau_2^2}{(1+\tau_2^2)^2}.
\end{equation}

 Frequency is normalized by the angular velocity $\Omega$ and the wavenumber $k$ is normalized by the disc scale height $H$, where for a thin disc we have $H=c_{\rm s}/\Omega$.   We also then confirmed that our dispersion relation reduces to that was obtained by \cite{Taka} if dust particles were single-size. If we set $\epsilon_2 =0$ or the dimensionless stopping time $\tau_2$ tends to infinity, the resulting growth rates are found numerically  consistent with results of  \cite{Taka} for a system with single-size particles (e.g., see figure 2).
\begin{figure}
\includegraphics[scale=.40]{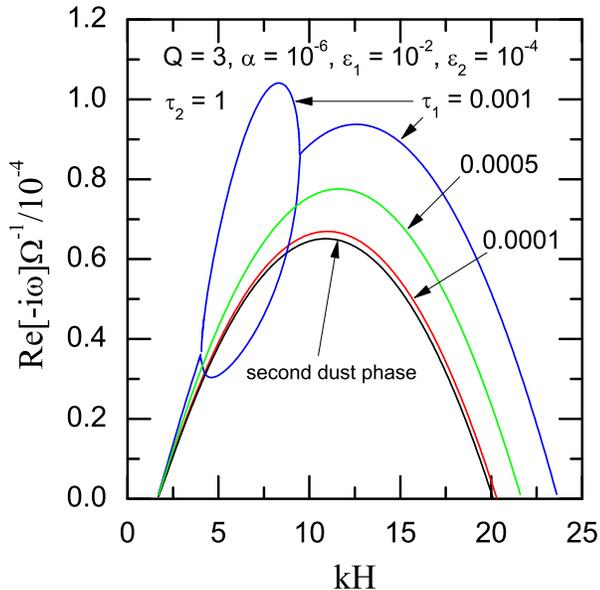}
\caption{ Normalized  frequency of the unstable modes, i.e. ${\rm Re}[-{\rm i}\omega]/10^{-4}\Omega $, as a function of the normalized wavenumber $kH$ for $Q=3$, $\alpha =10^{-6}$, $\epsilon_1 = 0.01$, $\epsilon_2 =0.0001$, $\tau_2 =1.0$ and different stopping time $\tau_1$. Each curve is labeled by the corresponding value of $\tau_1$. If the second dust component is neglected,  the first dust phase by itself is stable. But the second dust phase, in the absence of the first component, is unstable and the corresponding frequency is marked by "second dust phase".}
\end{figure}
\begin{figure}
\includegraphics[scale=.40]{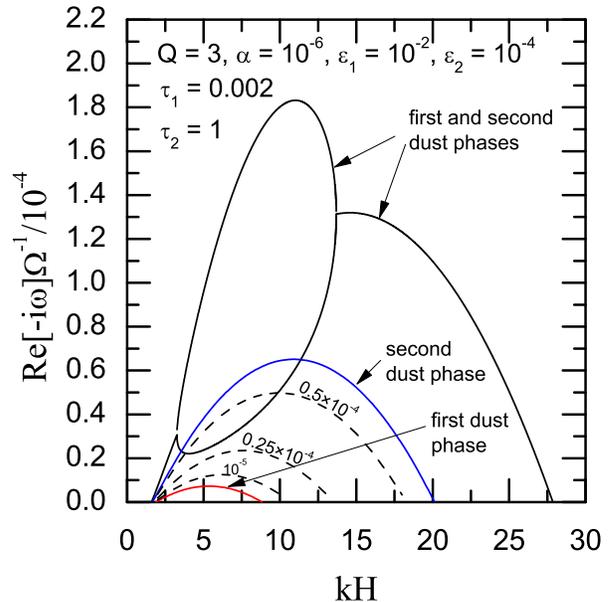}
\caption{ Same as Figure 1, but for $\tau_1=0.002$. Now, the first dust component is unstable in the absence of the second dust component and its curve is marked by  "first dust phase". As previous figure, the second component is unstable and its curve is marked by "second dust phase". Note that frequency of the instability for the "first dust phase" is much smaller than the "second dust phase" due to strong coupling of the first component to the gas. When both components are considered, frequency of the unstable modes significantly increases and for a range of the wavenumbers there are two unstable modes, though one of them is larger than the other one. We also consider cases with smaller values of $\epsilon_2$ which are shown by dashed lines. Each dashed curve is labeled by the corresponding values $\epsilon_2$, i.e. $0.5\times 10^{-4}$, $0.25\times 10^{-4}$ and $10^{-5}$. If $\epsilon_2$ becomes less than $10^{-5}$, then dispersion curve reduces to the single size analysis and the effect of second dust component on the instability becomes completely negligible.}
\end{figure}
\begin{figure}
\includegraphics[scale=.40]{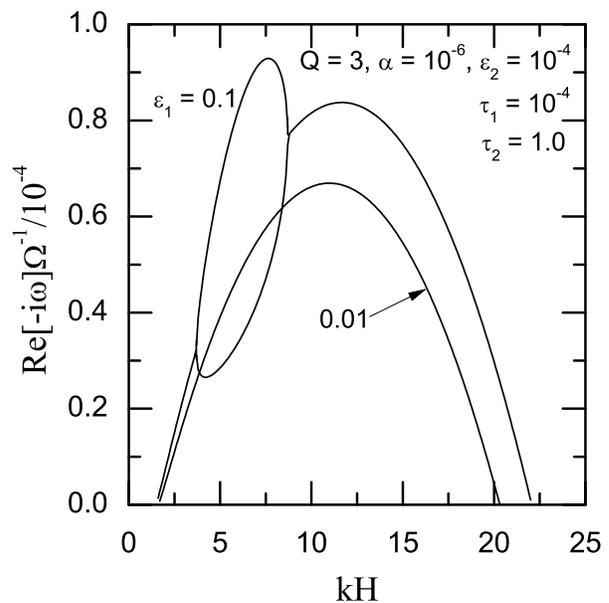}
\caption{ Normalized  frequency of the unstable modes, i.e. ${\rm Re}[-{\rm i}\omega]/10^{-4}\Omega $, as a function of the normalized wavenumer $kH$ for $Q=3$, $\alpha =10^{-6}$, $\tau_1 = 0.0001$, $\epsilon_2 =0.0001$, $\tau_2 =1.0$. Frequency of the unstable mode is shown for $\epsilon_1 =0.01$ and 0.1. As the mass of the first component increases, the instability occurs faster.}
\end{figure}
\begin{figure}
\includegraphics[scale=.40]{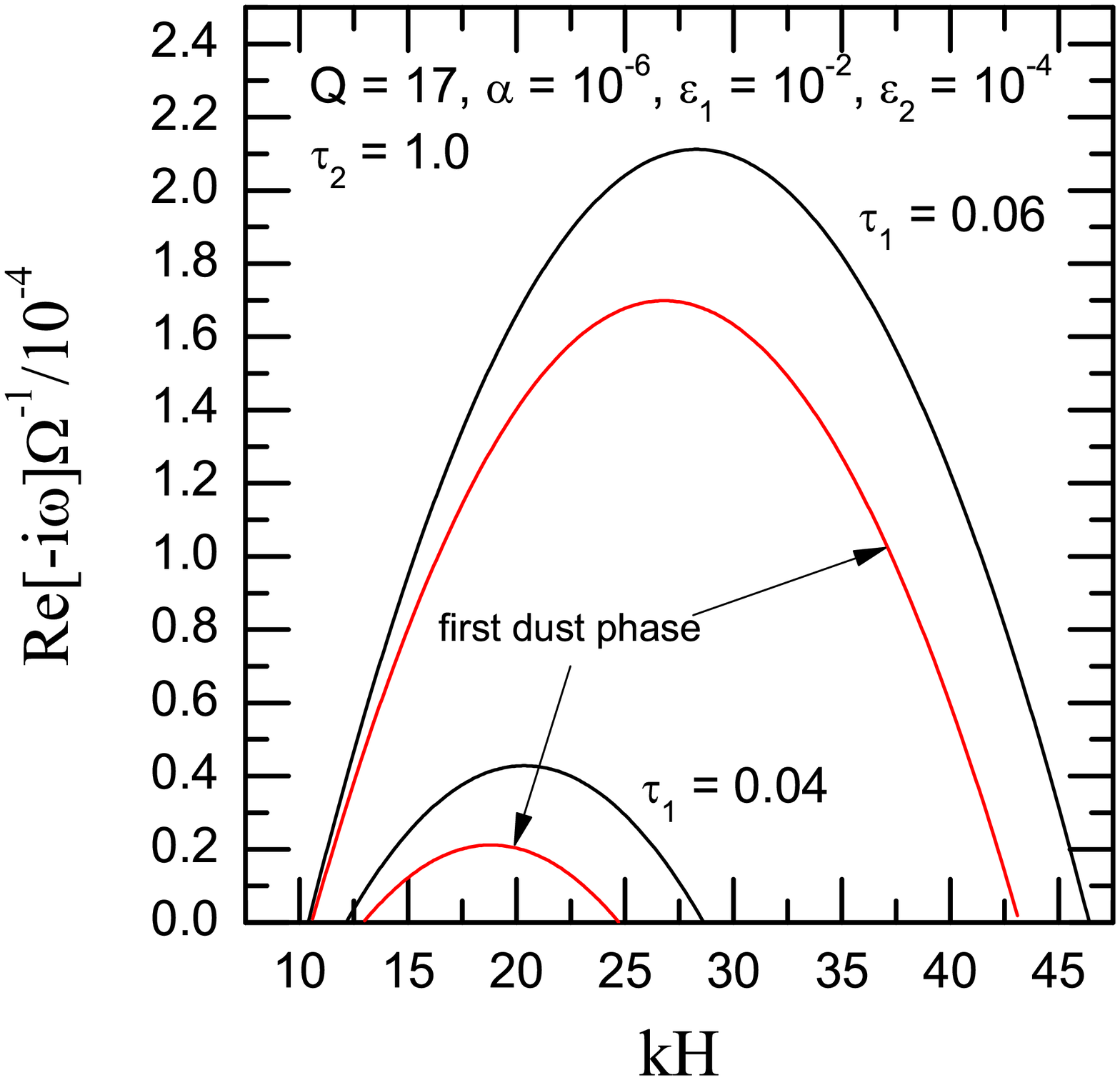}
\caption{ Normalized  frequency of the unstable modes, i.e. ${\rm Re}[-{\rm i}\omega]/10^{-4}\Omega $ as a function of the normalized wavenumer $kH$ in the minimum mass solar nebula at the radial distance 100 AU where the Toomre parameter is $Q=17$. Other input parameters are $\alpha =10^{-6}$, $\epsilon_1 = 0.01$, $\epsilon_2 =0.0001$, $\tau_2 =1$, and different values for the dimensionless stopping time $\tau_1$ are adopted, i.e. $0.04$ and $0.06$. We found that the system is stable if $\tau_1 < 0.04$. When the second dust component is neglected, the frequency of the unstable modes is shown and the corresponding curves are marked by "first dust phase".}
\end{figure}

In order to illustrate how the instability operates in a PPD  consisting of dust particles with two characteristic sizes, we consider the extreme case where the mass of the first dust phase is much larger than  the second dust phase. The first phase of  particles are well-coupled to the gas, however, the second dust phase is only marginally coupled. Such a configuration enables us to investigate how a tiny amount of large dust particles can modify the instability and affect dust clumping. We adopt the standard dust to gas ratio for the first phase which is about $\epsilon_1 = 0.01$ \citep{hayashi81}, but  a much smaller value for the dust to gas ratio of the second phase is used, i.e.  $\epsilon_2 = 0.0001$. So, the mass fraction of second dust phase  is smaller than the first dust phase by a factor of 100. Figure 1 shows  normalized  frequency of the unstable modes, i.e. ${\rm Re}[-{\rm i}\omega]/10^{-4}\Omega $ of the unstable mode as a function of the normalized wavenumber  $kH$ for $Q=3$, $\alpha =10^{-6}$, $\tau_2 =1$ and different values of $\tau_1$ which means coupling of the first dust phase varies from a well-coupled case  with $\tau_1 = 0.0001$ to a less coupled particles with $\tau_1 = 0.001$.  We found that if the second dust phase is neglected, the system is stable for this particular set of the input parameters. But if the first dust phase is not considered, the system becomes unstable and the corresponding frequency of the unstable mode is marked in Figure 1 by "second dust phase". It is found that the first dust phase is stable by itself (i.e., in the absence of second component) provided that the value of dimensionless stopping time $\tau_1$ is less than 0.001, but the second dust phase is unstable. Once the stopping time $\tau_1$ becomes slightly larger than 0.001, as we will show in the next figure, the first component becomes unstable in the absence of the second dust phase. 

Note that there are unstable modes despite the fact that Toomre parameter is greater than one which means the instability is driven by the drag force. Figure 1 reveals an interesting feature of the drag-driven instability because of the presence of larger dust particles. If this tiny amount of dust particles is {\it not} considered, linear analysis predicts that the system is stable so long as $\tau_1 \lesssim 0.001$. But presence of the second component leads to instability which its frequency is very close to the "second dust phase" for $\tau_1 \approxeq 0.0001$. But as the first dust component loses its coupling to the gas (i.e., larger $\tau_1$), the instability becomes more efficient and the frequency increases by even a factor of two. When $\tau_1$ is 0.001, we also found two unstable modes for a range of wavenumbers, though one mode is greater than the other one. In this figure, only those values of $\tau_1$ are adopted that imply a stable configuration for the "first dust phase" in the absence of the second component.

Figure 2 is similar to figure 1, but with a  value of $\tau_1$ which corresponds to {\it unstable} "first dust phase" in the absence of second component. Again, enhancement of the frequency is seen when both components are considered. Note that the increment of the frequency is a by factor of around three larger than "second dust phase", but comparing to the "first dust phase" this increment factor is much larger. Thus, presence of the second dust phase, despite of smallness of its mass in comparison to the first phase of particles, will significantly amplifies the instability.  We note that if the amount of large dust particles is completely negligible, the growth rate reduces to that of single size analysis by \cite{Taka}. In Figure 2, we also consider cases with smaller values of $\epsilon_2$ which are shown by dashed lines. While the enhancement of the growth rate for a case with $\epsilon_2 =10^{-4}$ is significant, the dispersion curve tends to a single size analysis for $\epsilon_2$ smaller than $10^{-5}$. This means a certain amount of large dust particles is necessary for significant enhancement of the growth rate. 

We can calculate ratio of the perturbed quantities such as $\delta\Sigma_{ d2}/\delta\Sigma_{ d1}$ and $\delta v_{2,x}/\delta v_{1,x}$ for the input parameters in figures 1 and 2. It is found that $\delta v_{2,x}/\delta v_{1,x}=4950,$ 1000, 500 and 250 for cases with $\tau_1 = 0.0001$, 0.0005, 0.001 and 0.002, respectively. This ratio of the perturbed radial velocities is independent of the wavenumber of perturbations and is much larger than unity which means mobility of the second dust phase is larger than the first dust phase. In fact, the second dust phase is marginally coupled to the gas and its movement due to the self-gravity of the system which is dominated by the first dust phase is easier. But the first dust phase is strongly coupled to the gas and so,  motion of the particles is under influence of the gas component which is gravitationally stable. We can see that as coupling of the first dust phase to the gas becomes weaker, the above ratio of the perturbed velocities reduces. For a case with $\tau_1 = 0.0001$, the ratio of the perturbed densities $\delta\Sigma_{d2}/\delta\Sigma_{d1}$  reduces from 46 to 34 as the normalized wavenumber increases. But as the coupling of the first dust phase to the background gas becomes weaker (i.e., $\tau_1$ increases), this ratio of the perturbed densities drops to much smaller values so that for cases with $\tau_1 = 0.001$ and 0.002, the ratio is found around 4 and 1.8, respectively. Thus, the amplitudes of the perturbed densities of the first and the second dust phases  becomes more or less comparable. In this case, concentration of the dust phases  due to SGI leads to significant dust clumping.

We can now consider dust clumping  due to this instability in the minimum mass solar nebula \citep{hayashi81} at a radius about 100 AU where observation evidences imply existence of ring-like structures. The surface density and the sound speed are power-law functions of the radial distance \citep{hayashi81}: $\Sigma (r) = 1.7\times 10^{3} (r/1 {\rm AU})^{-3/2}$ g${\rm cm}^{-2}$ and $c_{\rm s}(r)=1.0\times 10^{5} (r/1 {\rm AU})^{-1/4}$ cm${\rm s}^{-1}$.    As long as size of the particles is smaller than the mean free path of the gas, the gas drag force is in the Epstein's regime. The mean free path  is $\lambda = m_{\rm g}/(\sigma_{\rm mol} \rho_g )$ where $m_{\rm g}=3.9\times 10^{-24}$ g and $\sigma_{\rm mol} =2\times 10^{-15}$ cm$^{-2}$ is the collisional cross section of gas molecules and $\rho_g$ is the density of the gas. Thus, in the minimum solar mass nebula the mean free path becomes $\lambda = 1.15 (r/{\rm 1 AU})^{11/4}$ cm. In the minimum solar mass nebula for the distances larger than 1 AU from the central star, the mean free path of the gas is larger than 1 cm and at the radial distance $r=100$ AU we have $\lambda =3.6$ km. We are considering dust particles with sizes smaller than this value. Assuming that each dust particle with homogeneous internal density $\rho_m $ has a spherical shape with radius $a$, the nondimensional stopping time is $\tau = (\rho_m a / \rho_g v_{th})\Omega_{K}$, where   $v_{th}=(8/\pi)^{1/2}c_s $ is the mean thermal velocity. In this work, we adopt $\rho_m = 2 $ g ${\rm cm}^{-3}$ and so, the nondimensional stopping time at the midplane of the minimum mass solar nebula becomes $\tau = 1.8\times 10^{-7} (a/1\mu {\rm m})(r/{\rm 1 AU})^{3/2}$. Moreover, in the minimum solar mass nebula the Toomre parameter is written as $Q\simeq 56 (r/ {\rm 1 AU})^{-1/4}$ which implies $Q\simeq 17.7$ at $r=100$ AU. We now examine behavior of the instability for Toomre parameter $Q =17$, but the mass fraction of the first and the second dust phases are similar to figure 1, i.e. $\epsilon_1 =0.01$ and $\epsilon_2 = 0.0001$. As before it is assumed the second dust component is weakly coupled to the background gas flow, i.e. $\tau_2 =1$. However, we prefer to consider cases where the first dust phase is strongly coupled to the gas. We found that the system is stable as long as $\tau_1$ is approximately smaller than  0.04. In comparison to the previous figures, it is simply because a much larger value for the Toomre parameter is adopted here and the gas phase is gravitational stable. For $\tau_1 \lesssim 0.04$, the instability is not driven by the drag force because of its weakness comparing to the gravitational stability of the gas and coupling of the dust particle to the gas. For smaller Toomre parameter, this threshold value of $\tau_1$ occurs at smaller values as we found in the previous figures for $Q=3$. The threshold value $\tau_1=0.04$ at $r=100$ AU corresponds to dust particles with size 222 $\mu$m. Thus, particles with a size smaller than this size do not contribute to the instability.  Figure 4 shows frequency of the instability for $Q=17$ and different values of $\tau_1$. If we neglect the second dust component, the first dust phase is unstable and its frequency as a function of the wavenumber is shown here. But the second dust phase is found to be stable in the absence of the first component. Existence of the second dust component enhances the instability so that for $\tau_1 =0.04$, the fastest growing mode by a factor of around two is faster than a case without the second component.

 We also calculated the ratio of perturbed quantities such as $\delta\Sigma_{d2}/\delta\Sigma_{d1}$ and $\delta v_{2,x}/\delta v_{1,x}$ and it is found that $( \delta\Sigma_{d2}/\delta\Sigma_{d1},\delta v_{2,x}/\delta v_{1,x}) \simeq (0.07,12.5)$ and $(0.1, 8.36)$ for $\tau_1 =0.04$ and $0.06$, respectively. Thus, the second dust phase collapses harder  than the first dust phase because the amplitude of the perturbed density for the second dust phase is smaller than the first dust phase. However, radial velocity of the second dust phase is much larger than the first dust phase.

 Here, we have $\tau_2 =1$ which implies that the size of the second dust component at the radial distance 100 AU is $a=5.6$ mm. For $\tau_1 = 0.04$ and without the second component, the most unstable wavelength is about 18 AU and the corresponding growth time becomes 7.5 million years. But in the presence of a second dust phase, the growth time of the most unstable mode reduces to 3.7 million years at a wavelength around 15.8 AU. A case with $\tau_1 =0.06$ is examined in figure 4 which corresponds to particles with a size around 333 $\mu$m. Then, the most unstable mode occurs at wavelengths 12.6 AU and 11.9 AU corresponding to cases with only the first dust phase and including both components, respectively. Growth time of the most unstable mode for the first dust phase is $9.5\times 10^5$ years and this time reduces to $7.5\times 10^5$ years when the second dust component is considered.  We find that as coupling of the first dust phase becomes weaker, then the most unstable wavelength reduces and the growth timescale becomes faster. Thus, the resulting  dusty ring width varies depending on the size of first  dust phase.

 We explored a wide range of the input parameters and the results were similar to what we discussed so far. The number of the unstable modes were one or two depending on the input parameters. Because of  complexity of the dispersion equation and the number of the input parameters,  however, we could not find a closed analyticl condition for the stability of system. But the extreme cases that we studied clearly demonstrated that interactions between two species of dust via the background gas flow in the equations of motion and through their gravitational potential contributions in the Poisson equation are certainly important. Even in cases with two unstable modes none of them could be obtained independently from the single-size dust analysis as we showed (see previous figures). For example, in the extreme case that we considered in figure 1, single-size dust analysis predicts that the first dust component is stable in the absence of the second component. But our two-size dust analysis shows that the system is unstable due to existence of a second dust component and its interactions. 

Speeding up growth rate of the instability would ensure that the supply of accumulated mass can rise to meet the necessary conditions of having larger objects; otherwise these particles would drift toward   the central star before clumping. In order drag-driven instability to be a viable process, however, the growth time of the instability must be shorter than the radial drift time. Using this physical constraint, one can estimate the minimum required dust abundance to satisfy this condition. \cite{Takeuchi} estimated the minimum dust abundance must be less than 0.1 for SGI to operate at the midplane dust layer of a PPD. They considered two mechanisms for dust accretion, i.e. gas drag on individual particles and turbulent drag acting on the surface of the dust layer which both assist SGI, however, their analysis is restricted to single-size dust particles. In the light of our finding that even a slight amount of second dust phase enhances the efficiency of the drag-driven instability, we think, the minimum required mass would be still lower than the estimate of \cite{Takeuchi} when second dust phase is considered.

An interesting feature of the instability is the interval of wavelengths where the instability would operate.  Note that previous studies of SGI which neglect the gas dynamics predict that the short wavelength modes are stabilized by the dust diffusion. But  as has been shown by \cite{Taka}  when gas feedback is considered explicitly, long wavelengths are stable. We also confirm this finding when multiple grain species present. Interval of wavelengths for which there is instability becomes wider as   second dust phase coupling to the gas decreases. Our finding is also consistent with results of \cite{Laibe} who investigated streaming instability including two dust phases. They found that streaming instability becomes more efficient if two grain sizes present. 

 We can suppose that a PPD at its very early life consisting mostly of very small dust particles and a tiny amount of larger particles. As time passes, concentration of the small dust particles increases   due to a mechanism like drag-driven instability. Moreover, this process is accelerated in the presence of a  second population of dust particles. Since concentration of dust particles is enhanced due to the instability, probably it would be possible that small dust particles  merge to form larger particles. Under these circumstances, the system is still subject to the drag-driven instability and its efficiency  increases because of the emergence of multiple grain sizes. In other words, process of clumping due to drag-driven instability is accelerated. With increasing size of the particles, stopping time increases and the particles  gradually decouple from the gas. Once stopping time becomes larger than one, the streaming instability play the dominant role in clumping.

\section{Conclusion} 
One of the biggest challenges for theories of planet formation is how to explain accumulation of dust particles into larger objects. Among various mechanisms which have been proposed for planetesimal formation, SGI has a good efficiency for clumping of small dust particles. In contrary to the most of the previous analytical studies of this instability which assume all particles are single-size, however, we extend the analysis to situations with two grain sizes. We found this modification significantly assists growth of the instability. Contribution of even a small amount of the second dust phase to the instability is very remarkable especially when its coupling to the gas becomes weaker. Yet a more realistic analysis of drag-driven instability for a given grain size distribution is needed. 

\section*{Acknowledgments}
I am very grateful to the anonymous referee for his/her very useful comments and suggestions which greatly helped me to improve the paper. This work has been supported financially by Research Institute for Astronomy \& Astrophysics of Maragha (RIAAM) under research project No. 1/3720-58.

\bibliographystyle{apj}
\bibliography{reference}

\end{document}